\documentclass[prb,twocolumn]{revtex4}
\usepackage{graphicx}

\newcommand{\beq}{\begin{equation}}
\newcommand{\eeq}{\end{equation}}
\newcommand{\beqa}{\begin{eqnarray}}
\newcommand{\eeqa}{\end{eqnarray}}

\begin{document}

\title{Transport in disordered graphene nanoribbons} 

\author{Ivar Martin$^1$ and Ya.~M.~Blanter$^{2,3}$}
\affiliation{$^1$Theoretical Division, Los Alamos National
Laboratory, Los Alamos, New Mexico, 87544, USA\\
$^2$Kavli Institute of Nanoscience, Delft University of
Technology, Lorentzweg 1, 2628 CJ Delft, The Netherlands\\
$^3$Centre for Advanced Study, Drammensveien 78, 0271 Oslo,
Norway}

\date{\today}
\begin{abstract}
We study electronic transport in graphene nanoribbons with rough
edges.  We first consider a model of weak disorder that corresponds to
an armchair ribbon whose width randomly changes by a single unit cell
size.   
We find that in this case, the low-temperature conductivity is
governed by an effective 
one-dimensional hopping between segments of distinct band
structure.  We then provide numerical evidence and qualitative
arguments that similar behavior also occurs in the limit of strong
uncorrelated boundary disorder.

\end{abstract}
\maketitle

\section{Introduction}

Since the invention of the method for production graphene\cite{Geim},
many creative ideas for physical effects and devices have been put
forth\cite{review}. Whereas early papers emphasized unusual electron
properties of graphene as compared with ordinary metallic and
semiconductor materials, it had been soon realized that graphene
is a promising material for implementation of previously
known physical devices with considerably improved characteristics. One of
possible applications would be in semiconductor technology: 
excellent mechanical properties, easily tunable electron
concentration, zero nuclear spin, and simple production are among the
advantages that make graphene a much sought after material. However,
the drawback preventing 
the use of graphene for semiconducting applications is exactly what is
usually considered to be its main feature --- the absense of a gap
in the spectrum.  
In the absence of the gap, it is impossible to make even the simplest
conventional electronic devices.  For instance, a graphene p-n
junction does not rectify current, even though it has some other
interesting properties due to the Klein tunneling\cite{Klein}.
Similarly,  hybrid graphene -- normal metal systems are conducting for
arbitrary gate voltage applied to graphene \cite{NGN}. The only known
way to open a gap in monolayer
graphene is to use confined geometries --- graphene quantum 
dots\cite{dots} and graphene nanoribbons (GNR).

The electronic structure of ideal GNR is theoretically well
established.  It is very sensitive to the ribbon geometry,
{\em i.e.} orientation relative to the crystal axes and their
exact width \cite{dress,ezawa,netohall,fertig}.  Within a
tight-binding model with only nearest neighbor hopping, 
GNR with zig-zag edges have flat near-zero-energy bands of extended
edge states, while ribbons with armchair edges, depending on the
precise width, can be either metallic or semiconducting with the
gap inversely proportional to the GNR width. Numerical
studies~\cite{barone} show that passivation of the edges of ideal
GNR --- chemical bonding of edge carbon atoms with hydrogen ---
may open a small energy gap that does not scale with the GNR
width.

Recently, first experimental observations of transport in GNR have been
reported \cite{chen,kim,dai,ihn,lan}. GNR with widths in the range
from about 10 nm to 100 nm and lengths in the micrometer range have
been studied. The fabrication procedure does not yet allow to control
the GNR width with atomic precision (although chemical
fabrication\cite{dai} may eventually yield controlled edge
fabrication). As a result the edges are
disordered on the atomic length scale, as well as show longer-range
width variation of a few nanometers. For narrow enough ribbons ($ 
\lesssim 50$ nm) an unambiguous signature of the {\em geometric} gap $E_g$
scaling with the inverse {\em average} ribbon width has been extracted
from the gate voltage and temperature dependencies of
conductivity\cite{chen,kim}. In 
particular, in Ref.~{\onlinecite{chen}}, in a broad range of temperatures,
$T$, conductivity scales as $e^{-E_g/T}$.  The measured gap is a
smooth function of the ribbon width, and is insensitive to the GNR
orientation relative to the crystal axes.  Also, $1/f$ current noise
has been observed at low frequencies, $f < 100$ Hz, with the intensity
proportional to the GNR width \cite{chen}. 

These experimental results are definitely inconsistent with the theory for
ideal GNR that predicts different behavior depending on the orientation,
typically with many low-energy states. The observed effects are
clearly due to disorder. Indeed, it is natural to expect that any
disorder, bulk or boundary, should lead to Anderson localization and
open a transport gap; however, one would expect that this gap should
be defined by the strength of disorder, rather than by the GNR width.

In this work we provide a qualitative resolution to this apparent
puzzle 
by showing that electronic properties of the disordered GNR are indeed very
different from the clean GNR. We demonstrate that for the states near
the middle of 
the band, edge disorder leads to {\em segmentation} of the wavefunctions into
blocks of length of the order of GNR width. Thus, at low temperatures, the
system maps onto an effective one-dimensional (1D) hopping insulator
\cite{mott}.  We illustrate this behavior first with a model where disorder is
introduced through slowly fluctuating ribbon width, which allows more direct
numerical and analytical analysis (Section \ref{weak_disorder}). Then
we generalize the results to the experimentally relevant case of
strong disorder (Section \ref{strong_disorder}). Discussion and
conclusions are presented in Section \ref{conclusions}). 

\section{Weak disorder} \label{weak_disorder}

\begin{figure}[h]
\includegraphics[width=.9\columnwidth, height = 1.3cm]{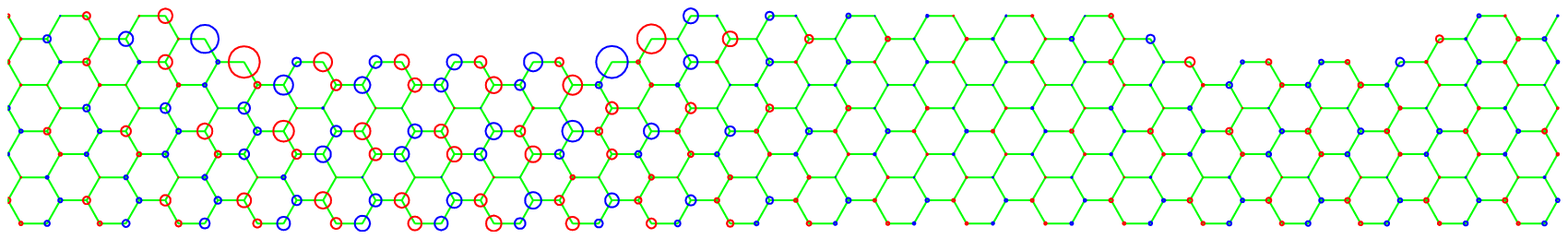}
\vspace{3 mm}

\includegraphics[width=.9\columnwidth, height = 1.3cm]{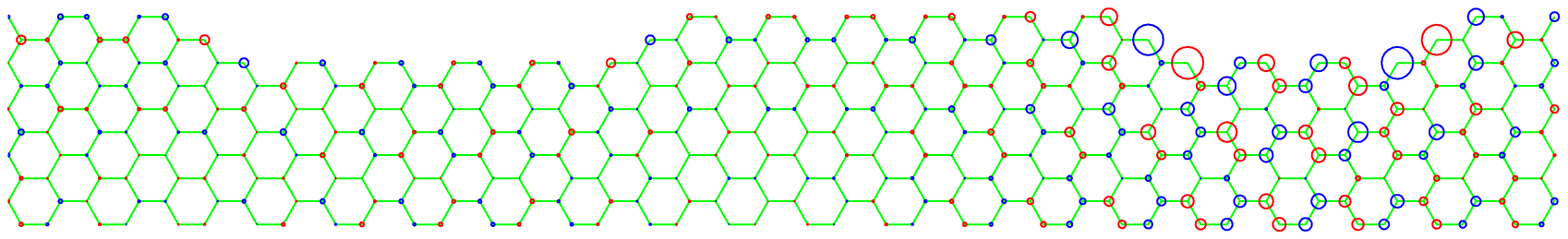}
\vspace{3 mm}

\includegraphics[width=.9\columnwidth, height = 1.3cm]{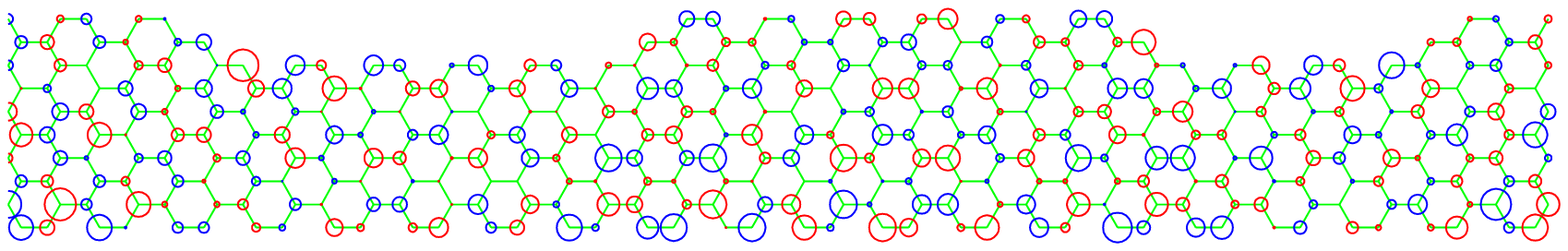}
\caption[] {Structure of the electronic wavefunctions in a ``weakly disordered"
armchair GNR. Periodic boundary conditions are applied in the horizontal
direction. In our convention, the narrow segments have width $W = 4\sqrt{3}$,
and in the limit of infinite length would have no band gap; the surrounding
regions (width $W = 5\sqrt{3}$) for an infinitely long ribbon would have a gap
$E_g = 2\times 0.169t$ around zero energy (the middle of the GNR $\pi$ band).
The radii of the circles are proportional to the site amplitudes of the
wavefunction, with the color representing sign. The top plot corresponds to a
the lowest energy state inside the gap, $E = -0.096t$, spatially localized in
the left (longer) metallic segment; the middle -- to the low energy state in
the right (shorter) metallic segment, $E = -0.131t$; the bottom is a
delocalized state well outside the gap, $E = -0.497t$. Note the abrupt change
in the localized wavefunctions' amplitude at the ``interface," and rather
uniform amplitude across the ribbon. \label{fig:wd_wf}}
\end{figure}

Let us consider an armchair GNR.  An ideal ribbon of the width $W$, measured in
units of minimal carbon-carbon distance, $a_g$, is metallic (no gap) for $W =
(3N+1)\sqrt{3}$, and semiconducting (with gap $E_g\sim t/ W$) for $W =
3\sqrt{3}N$ and $W = (3N+2)\sqrt{3}$, Ref.~\onlinecite{fertig}.  Here, $t$ is the
graphene nearest-neighbor hopping matrix element (we neglect the next-nearest
neighbor hopping which causes slight particle-hole asymmetry), and $N$ is an
integer. {\em Weak disorder} can be introduced as geometric fluctuations of the
ribbon width, such that the ``disordered" ribbon is comprised of ideal
segments of random length of order $L$, with width changing from
segment to segment.  We assume that $L 
> W$.  An example of a ``disordered" configuration of this kind is shown in
Fig.~\ref{fig:wd_wf}. While this situation has not been yet
realized experimentally, it has the advantage that its analysis is
straightforward, and, as we will argue, the behavior is
related to the experimentally relevant case of strong disorder. If
the length of each segment is longer than its width, to the lowest
order, one can consider individual band structure of each segment
separately. Depending on the width, some of the segments are
nearly metallic, with the finite size gap of about $t/L$, while
others are ``insulating'' with the gap $t/W$.  In
Fig.~\ref{fig:wd_wf} we show the results of numerical
diagonalization of the tight-binding graphene Hamiltonian ---
representative wavefunctions in different regions. Note that
although the segmentation is caused by the surface defect (change
of the width by just one row of atoms!), the wavefunctions show
high degree of uniformity {\em across} the ribbon, and rather
sharp confinement to the respective regions {\em along} the
ribbon. Thus, at low energies ($|E| < t/W$) it is natural to
represent the system by a one-dimensional hopping model,
\beq\label{eq:H} H = \sum_{i\alpha}{\epsilon_i^\alpha
\hat{c}_i^{\alpha\dagger} \hat{c}_i^\alpha} + \sum_{i\alpha,
j\beta}{t_{ij}^{\alpha\beta} \hat{c}_i^{\alpha\dagger}
\hat{c}_j^\beta} + h.c. \ . \eeq Here, the operator
$\hat{c}_i^{\alpha\dagger}$ ($\hat{c}_i^{\alpha}$) creates
(destroys) electron in the metallic ``grain" $i$ in the orbital
$\alpha$.

\begin{figure}[ht]
\includegraphics[width=.8\columnwidth]{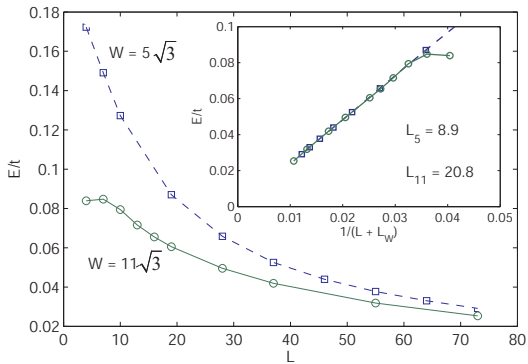}
\caption[]{The lowest energy state in a GNR as a function of
length $L$ of a ``metallic" constriction surrounded by an
``insulator."  The configuration is similar to the one in
Fig.~\ref{fig:wd_wf}.  The total length of the ribbon used in
simulation is $120a_g$ (with periodic boundary conditions along
horizontal axis).  The results presented are for two GNR widths,
$W = 5\sqrt{3}$ and $W = 11\sqrt{3}$,  with  $W = 4\sqrt{3}$ and
$W = 10\sqrt{3}$, respectively, in the metallic regions. In the
inset we fit the energy to the form $E \propto (L + L_W)^{-1}$.
The offset $L_W$ appears due to the leakage of the wavefunction
from the metallic regions into the insulating ones. As expected
(see text), $L_W \sim W$. \label{fig:E_L}}
\end{figure}

To complete the formulation of the effective model
Eq.~(\ref{eq:H}) we need to determine the distributions of the
on-site energies $\epsilon_i^\alpha$ and inter-site hopping matrix
elements $t_{ij}^{\alpha\beta}$.  For simplicity we assume that
the average length of the segments, both insulating and metallic,
is the same, $L_{av}$.  The low-energy spectrum in the metallic
segments follows from the Dirac dispersion of the infinite
metallic armchair GNR \cite{fertig}, $\epsilon = c|k|$, where $c =
3ta_g/2$ and $k$ is the momentum along GNR.  The levels in a given
metallic segment of length $L$ are therefore approximately
equidistant, with the average level spacing $\sim t/L$. In
Figure~\ref{fig:E_L} we show the result of a tight-binding
calculation for the lowest energy state as a function of the
length of a metallic segment embedded in the insulating GNR.
Indeed we find that the energy scales approximately as $1/L$. Even
better fit is obtained by using the form $1/(L+L_W)$ which takes
into account the leakage of the wavefunction from the metallic
regions into surrounding insulating ones. Note that $L_W \approx
W$.  If the lengths $L_i$ for all grains were equal, the level
structures in all grains would be identical (apart from the small
splitting caused by inter-grain tunneling). However, for a
distribution of lengths, the energy levels in different grains are
likely to be out of registry by the amount $\sim t/L_{av}$.

We now evaluate the tunneling matrix elements $t_{ij}$ between low-energy
states in metallic segments. Tunneling occurs through the intermediate states
in the insulating regions. The states just outside the gap are particularly
important for tunneling.  Near the gap edge the dispersion is quadratic,
$\epsilon = \sqrt{c^2k^2 + (E_g/2)^2}\approx E_g/2 + c^2k^2/E_g$. This
corresponds to the effective mass in the insulating regions $m^*\sim (W a_g^2
t)^{-1}$. The tunneling amplitude through a barrier of height $E_g$ and length
$D$ can be estimated using the WKB approximation as $e^{-\alpha D/W}$, where
$\alpha$ is a numerical coefficient of order 1.  We have also verified this by
a direct tight-binding calculation of the tunnel splitting of energy levels in
two identical metallic segments separated by an insulating segment of variable
length.

From the distributions of $\epsilon_i^\alpha$ and $t_{ij}^{\alpha\beta}$, it
follows that for $L_{av} > W$, the level spacing in the metallic grains is
larger than the tunneling amplitude between the neighbors, making it impossible
to have metallic, i.e. band, conduction. The system is a one-dimensional
example of a simple impurity band insulator, a standard model used to describe
lightly doped compensated semiconductors \cite{efros_shklovskii}. The
finite-temperature conductivity of such insulator can be evaluated by standard
techniques \cite{mott},
\beq \sigma \sim e^{- \alpha n
L_{av}/W - t/(n L_{av}T)},
\eeq
where $T$ is the temperature, and $n$ is the length of the optimal jump.  By
minimizing the exponent, we find that $n_{opt} = \sqrt{t W/(\alpha L_{av}^2
T)}$. Hence there is a crossover from the nearest neighbor ($n_{opt} = 1$, NNH)
to variable range hopping ($n_{opt} >1$, VRH) at temperature $T^* \sim t
W/L_{av}^2$,
\beqa \sigma\sim \left\{\begin{array}{ll}
  e^{-2\sqrt{\alpha t/(WT)}} &\quad {\rm for}\ T < T^* \\
  e^{-\alpha L_{av}/W - t/(L_{av}T)} &\quad {\rm for}\ T^*< T< t/L_{av} \\
\end{array}\right.\label{eq:sigma}
\eeqa
Note that $T^* < t/W$, and thus both behaviors are possible within our model.
At temperatures higher that $t/L_{av}$ multiple states in the metallic regions
have to be included.  We do not consider here other regimes of one-dimensional
hopping \cite{Raikh} that can become relevant at very low temperatures.

\section{Strong disorder} \label{strong_disorder}

We now turn to the {\em strong} disorder case, when the boundary is randomized
at the atomic scale (this models the situation when some atoms are cut out or
replaced by other atoms, e.g. oxygens,  in the process of
fabrication).   Yet, we assume that the relative variation of the
ribbon width introduced by disorder is small.  This is different from
the near-granular case considered in Ref.~\onlinecite{sols}.   
 
\begin{figure}[htb]
% good files end with _bb, however, on arxive they do not show, so had
% to replace them with
% lower quality bbi2 but with correct bounding box

\includegraphics[width=.9\columnwidth, height = 1.7cm]{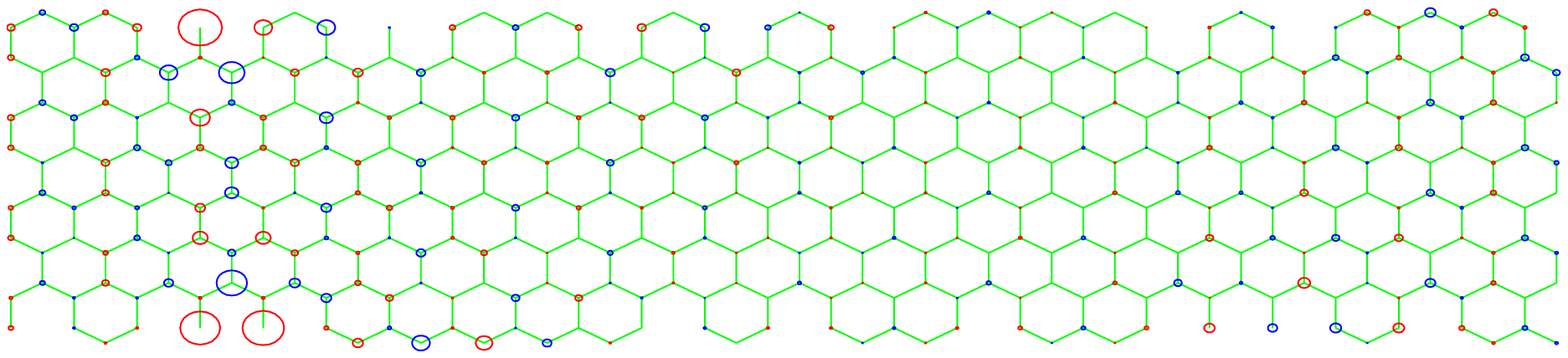}
\vspace{3 mm}

\includegraphics[width=.9\columnwidth, height = 1.7cm]{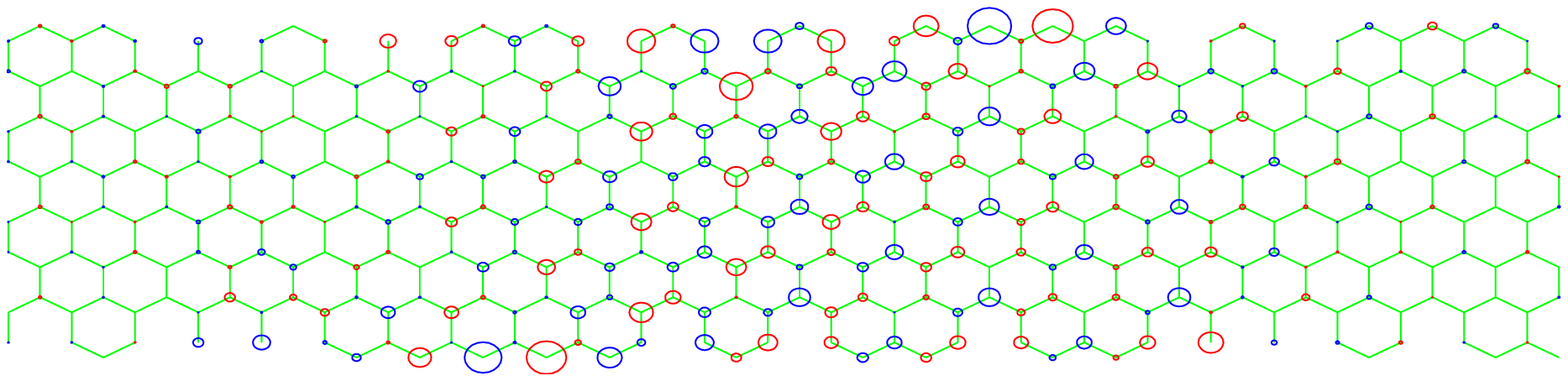}
\vspace{3 mm}

\includegraphics[width=.9\columnwidth, height = 1.7cm]{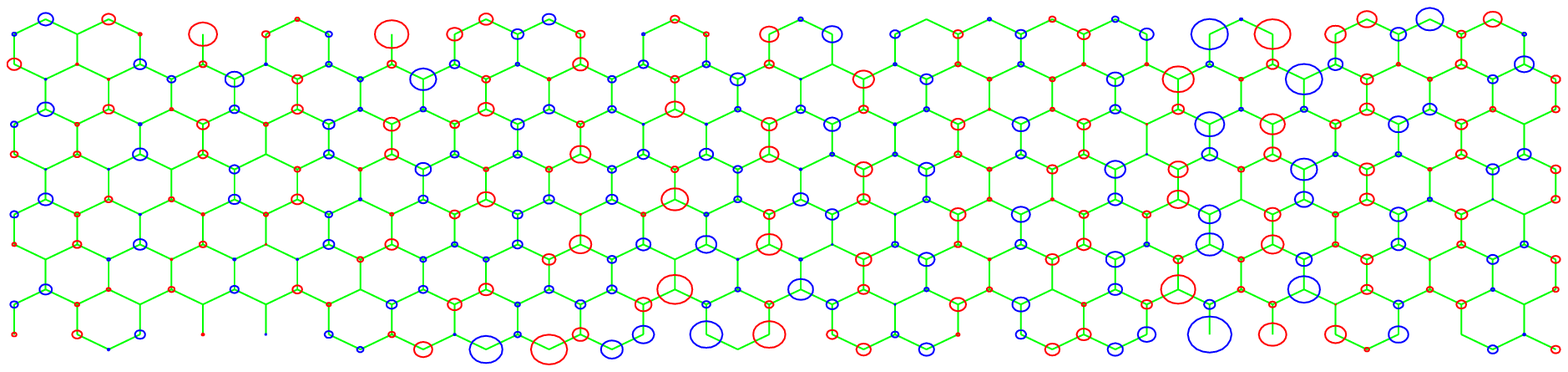}
\caption[]{Structure of the electronic wavefunctions in a strongly disordered
zig-zag GNR. The disorder is generated by  randomly eliminating half of carbon
atoms at the edges of GNR.  Periodic boundary conditions are applied in the
horizontal direction.  The energies of the states, from top down, $E =
-0.071t,\ -0.089t,\ -0.255t$.   Note how the confinement length increases away
from the center of the band ($E = 0$).  The typical confinement length at the
energies inside the gap is of the order of the ribbon width.
\label{fig:wd_wf2}}\vspace{-4 mm}
\end{figure}

An example of a ``strongly" disordered configuration that corresponds
to small relative variation in the ribbon width is shown in
Fig.~\ref{fig:wd_wf2}.  We chose a perfect zig-zag nanoribbon as the
reference structure. Ideal zig-zag nanoribbons are always 
metallic, for any ribbon width, due to the presence of the edge 
states~\cite{dress}. We observe that edge disorder (here generated by
eliminating at random half of the sites along the edges) leads to the
wavefunction localization.  However, since disorder is now short-correlated,
the wavefunctions no longer have a typical extent along the ribbon, but rather
can be either more or less localized.  We find numerically that the
wavefunctions corresponding to the low-energy states ($|E|< t/W$) that are
highly localized along the direction of the boundary ($L \ll W$, {\em e.g.}
Fig.~\ref{fig:wd_wf2}a) also do not penetrate deep inside the ribbon, having
large amplitude only near the surface. On the other hand, states that are more
extended along the ribbon also penetrate deeper into the bulk.  This effect can
be traced back to the behavior of the edge states in zig-zag GNR -- the
wavevector along the ribbon for these states is approximately equal to their
exponential decay length into the bulk.  In effect, in the absence of
next-nearest-neighbor hopping, one can have states with very low
energy, $\sim t e^{-W/a}$, localized over the distance of about single unit
cell near the ribbon edge.  The number of such states within a segment
of length $W$ can be easily estimated to be $|K - K'|/(2\pi /a W) \sim
W$, that is, each boundary atom can support one highly localized low
energy state.  Their spectrum which can be derived from the dispersion of
the ideal ribbon surface states is \cite{fertig} $E \sim t\exp(-kW)$
for $kW \gg 1$, where $k$ is the wave vector deviation from the Dirac
point. This spectrum provides the density of states which behaves as
$E^{-1}$ at $E \ll t/W$. 

This extremely high low-energy density of states is obviously an artefact of our
model and disappears 
if the hopping between next-nearest neighbors (NNN) on the graphene
lattice is taken into account. It is characterized by
the overlap intergral $t'$ which in graphene approximately equals to
$0.2t$.  In the following, we assume $t' > t/W$ which is the case for
all graphene nanoribbons studied in the experiments. If the NNN overlap
is taken into account, the edge states are hybridized and form a band
of the width $\sim t'$. Thus, the (one-dimensional) density of states
in the gap $\vert E 
\vert < t/W$ is approximately constant and equal to one state per
surface atom per $t'$.  Since the localization length is governed by
the energy distance to the next subband, it is nearly independent of
$t'$ and still is approximately equal to $W$.  We therefore
coarse-grain over the ribbon elements of size $W\times W$ to obtain
the low-energy level spacing within a coarse-grained element $\Delta E
\sim t'/W$.  
Now, we can formulate the variable range hopping conductivity between
these elements as  
\beq \sigma \sim e^{- \alpha n - t'/(n W T)}.
\eeq
Optimizing over $n$, we find two regimes,
\beqa \sigma\sim \left\{\begin{array}{ll}
  e^{-2\sqrt{\alpha t'/WT}} &\quad {\rm for}\ T < T_c \\
  e^{- t'/(WT)} &\quad {\rm for}\ T_c < T  <  t/W \\
\end{array}\ ,\right.\label{eq:sigma2}
\eeqa
with the crossover temperature $T_c = t'/W <
t/W$. Again, higher-temperature (top line) and lower-temperature
(bottom line) ranges correspond to VRH and NNH, respectively.  

On the experimental side, Chen {\em et al.} \cite{chen} find that, {\em e.g.} in
20 nm wide GNR, which has a gap 28 mV, at relatively high temperatures, between
50 K and 100 K, transport is activated, $\sigma \propto e^{-E_g/T}$. The data
is lacking at intermediate temperatures; however, the single low-temperature
data point at 4 K shows conductivity much higher than would be expected from
simple activated hopping. This may reflect a crossover from the nearest to
variable range hopping. Our estimate for the crossover temperature is
about 5 Kelvin, which is consistent with the experimental results.
More detailed experimental data in the
intermediate temperature regime should allow direct test of our predictions.
It has also been found that the experimental size of the gap is
smaller than $t/W$, which may be consistent with our value $t'/W$.

\section{Discussion and conclusions} \label{conclusions}

The alternative explanation of experimental results was proposed in a
model respresenting a GNR with {\em very strong} interface disorder as
a chain of quantum dots, hosting localized electron states\cite{sols}.
It was suggested by the authors that the experimentally observed gap
has to do with charging energy of the ``quantum dots." Indeed, recent
experiments 
\cite{ihn,lan} indicate that there are local charging centers in
graphene nanoribbons.  Compared to the case studied in
Ref. \onlinecite{sols}, both cases that we considered above correspond to at
most {\em mild} disorder.  We therefore address now how the Coulomb
interaction will affect our results. 

The Coulomb interaction can modify low-temperature hopping
conductivity.  Interaction leads to opening of the soft Coulomb gap around the
Fermi surface, which enters as an energy cost inversely proportional to the
length of the hop \cite{efros_shklovskii}. Thus in the presence of the Coulomb
interaction,  the
expression for conductivity has to be modified as
\beq \sigma \sim e^{- \alpha n
- t'/(n W T) - e^2 /(\epsilon n a_g W T)}.
\eeq
Since $e^2/a_g t \sim 1$ in graphene, the Coulomb cost will become
relevant if the dielectric constant of the embedding medium $\epsilon$
is smaller than $t/t'\sim  5$.  While the functional form of
conductivity in this case remans the same as in Eq.~(\ref{eq:sigma2}),
the energy scale that defines the gap is different, $t'/W \rightarrow
e^2/(\epsilon a_g W)$. 

Thus, for freely suspended graphene the transport is in fact expected
to be dominated by the {\em soft} Coulomb blockade.  On the other
hand, placing graphene in the vicinity of high-$\epsilon$ medium or
metallic gate would reduce the Coulomb interaction strength and
range~\cite{BR}, leading to crossover to Mott's VRH. 

Finally, we note that the $1/f$ noise observed by Chen {\em et
al.} \cite{chen} may also be consistent with the scenario presented
here, that is, it may be {\em intrinsic}, rather than caused by
the charge fluctuations in the substrate, as was suggested in
Ref. \onlinecite{chen}.  Due to the presence of an exponentially broad
distribution of the tunneling rates in the hopping transport, the
experimentally observed Hooge relation \cite{hooge} between the
low-frequency current noise and the DC current, $I_\omega^2/I^2 =
A(\omega,T)/\omega$, can be naturally expected \cite{shklov}.
Straightforward application of the Shklovskii's argument
\cite{shklov} to one dimension leads to Hooge's parameter $A
\propto \exp(- B T)$ in the low-temperature (VRH) regime, and
approximately constant $A$ at high temperatures (NNH). Whether the
$1/f$ noise is indeed intrinsic can be tested by varying the
substrate properties, or performing measurement on a suspended GNR
\cite{geim2}.

We acknowledge useful discussions with M. Fogler and A. F. Morpurgo. IM
acknowledges the hospitality of Delft University of Technology,
where part of this work was performed. This work was supported by
EC FP6 funding (contract no. FP6-2004-IST-003673). Partial support
was provided by US DOE.

\end{document}